\documentclass[12pt,twoside]{article}
\usepackage{amsmath}
\usepackage{amssymb}
\usepackage{latexsym}
\usepackage{graphicx}

\newcommand{\xdot}{\dot{X}}

\newcommand{\xdddot}{\ddot{X}}
\newcommand{\tbr}{(t, {\bf r})}
\newcommand{\vvr}{({\bf v} \rho)}
\newcommand{\trrr}{\mathop{\mathrm{tr}}}
\newcommand{\Cpr}{Clebsch pa\-ra\-me\-ter\-iza\-tion}
\newcommand{\vA}{{\bf A}}
\newcommand{\vB}{{\bf B}}

\newcommand{\vE}{{\bf E}}

\newcommand{\vX}{{\bf X}}
\newcommand{\vv}{{\bf v}}
\newcommand{\vJ}{{\bf J}}
\newcommand{\vj}{{\bf j}}

\def\beq{\begin{equation}}
\def\eeq{\end{equation}}
\def\beqar{\begin{eqnarray}}
\def\eeqar{\end{eqnarray}}
\def\del {{\partial}}
\def\d {{\partial}}
\def \La {{\cal L}}

\let\vec\boldsymbol
\numberwithin{equation}{section}

\begin{document}
\title{Inserting Group Variables into Fluid Mechanics\footnote{XXV Group Theory Meeting, Cocoyoc, Mexico, August 2004.}}

\author{R. Jackiw\footnote{jackiw@lns.mit.edu}\\
\small\it Department of Physics\\
\small\it Massachusetts Institute of Technology\\
 \small\it Cambridge, MA 02139}
\date{\small MIT-CTP-3553}


\maketitle
\abstract{
A fluid, like a quark-gluon plasma, may possess degrees of freedom indexed by a group variable, which retains its identity even in the fluid/continuum description. Conventional Eulerian fluid mechanics is extended to encompass this possibility.}

\section{Introduction}
The dynamics of fluids is described by a classical field theory, whose origins lie in the nineteenth century, like Maxwell electrodynamics with which it shares some antecedents. The electromagnetic theory has enjoyed much development: it passed into quantum physics, and the resulting quantum electrodynamics served as a model for its non-Abelian generalization, Yang-Mills theory, which today is at the center of a quantum field theoretic description for fundamental physics. We believe that fluid dynamics can undergo a similar evolution and thereby be generalized to fluid systems with non-Abelian charges.

The quintessential example of a physical system with non-Abelian charges
is the quark-gluon plasma.
High energy collisions of heavy nuclei can produce a plasma state of quarks and gluons.
This new state of matter has recently been of great interest both theoretically
and in experiments at the RHIC facility and at CERN.
In fact, there is growing evidence that such a state has already been
achieved at
the RHIC facility \cite{mclerran}.

In attempting a theoretical description, there are basically two approaches that we can use. Since the plasma is at high temperatures, one can argue that the average energy per particle is high enough to justify the use of perturbative Quantum Chromodynamics by virtue of asymptotic freedom. However, it is known that because of infrared divergences various resummations, such as summing hard thermal loop contributions, have to be done before a perturbative expansion with control of the infrared degrees of freedom can be set up \cite{pisarski}. One has to address also the question of chromomagnetic screening, because unlike the Abelian plasma, there can be spatial screening of magnetic type interactions \cite{GPY}. The expected end result is then a good description valid at high temperatures and for plasma states that are not too far from equilibrium, since one is perturbing around the equilibrium state. An alternative approach, which may be more suitable for nondilute plasmas or for situations far from equilibrium, would be to use a fluid mechanical description.

We begin by recalling the conventional formulation of Eulerian fluid mechanics (in the Abelian case), for perfect, isentropic fluids \cite{bib4} The variables are the matter density field $\rho \tbr$ and the velocity field $\vv \tbr$; the two are united in the current density $\vj = \vv \rho$.

The equations satisfied by these in the non-relativistic situation are the continuity equation
\begin{equation}
\frac{\partial}{\partial t} \, \rho \, + \vec\nabla \cdot \vvr = 0 = \frac{\partial}{\partial t} \, \rho + \vec\nabla \cdot \vj,
\label{eqone}
\end{equation}
and the Euler equation,
\begin{equation}
\frac{\partial}{\partial t}\, {\bf v} + \vv \cdot \vec\nabla \vv =\,  \mbox{force density}.
\label{eqtwo}
\end{equation}
The first expresses fluid matter conservation; the second describes forces. The force density on the right could be derived from a pressure function, $P(\rho)$,
\begin{subequations}\label{eqthreesub}
\begin{equation}
\text{force density} = - \frac{1}{\rho}\, \vec\nabla P (\rho),
\label{eqthreea}
\end{equation}
or more generally, from an enthalpy function $V(\rho)$,
\begin{equation}
\text{force density} = - \vec\nabla V^{\prime} (\rho),
\label{eqthreeb}
\end{equation}
or in case of magneto hydrodynamics from the Lorentz force law in the presence of electric $\vE$ and magnetic $\vB$ fields
\begin{equation}
\text{force density} = e (\vE + \vv \times \vB).
\label{eqthreec}
\end{equation}
\end{subequations}
(Here $e$ is the charge; the velocity of light is scaled to unity; a dash, as in (\ref{eqthreeb}), denotes derivative with respect to argument.)
In the first two cases it is assumed that the dependence of $P$ or $V$ on $\rho$ is known ($=$ equation of state) and the system (\ref{eqone}), (\ref{eqtwo}) is self-contained. In the magneto hydrodynamical case, the the electromagnetic fields are taken to be either  prescribed functions of $\tbr$, or  determined self-consistently from the Maxwell equations with sources $(\rho, \vj)$.

A simplification arises when the vorticity $\vec\omega \equiv \vec\nabla \times \vv$ vanishes. Then the velocity can be written as the gradient of a velocity potential $\theta$
\begin{equation}
\vv = \vec\nabla \theta
\label{eqfour}
\end{equation}
and equations (\ref{eqtwo}), (\ref{eqthreeb}) are integrated once to yield the Bernoulli equation
\begin{equation}
\frac{\partial}{\partial t} \, \theta + \frac{1}{2} \, {\bf v}^2 = - V^{\prime} (\rho)
\label{eqfive}
\end{equation}

In the relativistic case, the matter density and current density are united in a current 4-vector $j^\mu = (\rho, \vj) = (\rho, \rho \vv)$ which is decomposed into proper density $n$ and  a proper velocity $u^\mu$, normalized to unity.
\begin{equation}
j^\mu \equiv n u^\mu \quad u^\mu u_\mu =1
\label{eqsix}
\end{equation}
This decomposition, which entails no restriction -- it is valid by definition -- is called the Eckart factorization \cite{bib5}. [Note that the Eckart factorization also holds non-relativistically $(\rho, \vj) = \rho(1, \vv)$, but the velocity is not normalized.]
The current is conserved, satisfying a continuity equation,
\begin{equation}
\partial_\mu j^\mu = 0,
\label{eqseven}
\end{equation}
and the Euler force equation is presented with the help of the Eckart factorization as an equation for the proper velocity $u$.
\begin{equation}
u^\mu \partial_\mu\ u^\nu = \text{force terms}
\label{eqeight}
\end{equation}

Equivalently, the non-relativistic/relativistic Euler equations can be presented as continuity equations for a non-relativistic/relativistic energy-momentum tensor $\theta^{\mu \nu}$.
\begin{equation}
\partial_\mu \theta^{\mu \nu} = 0
\label{eqnine}
\end{equation}
Additionally there exists a phase space formulation, with known brackets for $\rho$ and $\vv$, and a Hamiltonian generates the equations of motion by bracketing \cite{bib4}. Also there is a configuration space formulation in terms of a Lagrangian. This requires writing the velocity in terms of the \Cpr \ \cite{bib6}.
\begin{equation}
\vv = \vec\nabla {\theta} + {\alpha} \vec\nabla \beta
\label{eqten}
\end{equation}
Note that in three dimensions the $(\rho, \vv)$ phase space is 4-dimensional, hence even.
Also the velocity vector field $\bf v$ involves three functions, which are captured by the three scalar Monge potentials $(\theta, \alpha, \beta)$.  [That a Lagrangian/configuration-space formulation requires presenting the velocity vector field in terms of potentials, which are not seen in the equations of motion is analogous to the Lagrangian/configuration-space formulation for a charged particle moving in a magnetic field: the Lagrangian involves a vector potential, while the equation of motion sees only its curl -- the magnetic field.]

The research problem that my collaborators and I set for ourselves concerns the question of how to generalize the above, when there is a variety of fluids, with each species labeled by a Lie group index \cite{bib4}, \cite{bib7}. Certainly we expect that the current will inherit the group index $j^\mu \to J^\mu_a$, and will be group-covariantly conserved.
\begin{equation}
\partial_\mu J^\mu_a + f_{abc} \, A^b_\mu J^\mu_c \equiv (D_\mu J^\mu)_a
\label{eqeleven}
\end{equation}
Here $A^a_\mu$ is a non-Abelian gauge potential, describing an external or dynamical gauge field in which the fluid moves. In the latter case we expect that the gauge field obeys its own equation of motion.
\begin{equation}
\partial_\mu F^{\mu \nu}_a + f_{abc} \, A^b_\mu\, F^{\mu \nu}_{c} = (D_\mu\, F^{\mu \nu})_a = J^\nu_a
\label{eqtwelve}
\end{equation}
The $f_{abc}$ are structure constants for the appropriate group.

But the question that still remains is how should the density and the velocity of the fluid be related to the group covariant current; {\it viz}.\,what are the group transformation properties of the density and velocity; what is the analog of the Eckart factorization (\ref{eqsix}); which is the non-Abelian Euler equation that supplements the non-Abelian current conservation equation (\ref{eqeleven})?

\section{Non-Abelian Euler variables}\label{abeEul}
A possible form for the non-Abelian Euler fluid variables may be inferred
from the single-particle equations of motion --- the Wong equations \cite{bib8} (over-dot denotes time differentiation), 
\begin{subequations}\label{na1}
\begin{eqnarray}
m\, \xdot^\mu &=&
P^\mu, \label{na1a} \\
m\, \dot{P}_\mu &=& g{\cal Q}_a F^a_{\mu\nu}P^\nu, \label{na1b}\\
m\, \dot{\cal Q}_a &=& -g f_{abc} P^\mu A^b_\mu {\cal Q}_c,\label{na1c}
\end{eqnarray}
\end{subequations}
with $A^a_\mu$ and $F^a_{\mu\nu}$ describing a (background or dynamical) gauge field interacting with strength $g$. [We have assumed that the only active forces are Lorentz forces due to the gauge fields. Of course other forces can be included on the right of (\ref{na1b}).]
The equation of motion (\ref{na1c}) satisfied by the charge ${\cal Q}_a$ is postulated to hold so that the single particle current
\begin{equation}
J^\mu_a (t, {\bf r} ) = \int d\tau ~ {\cal Q}_a(\tau ) ~
\xdot^\mu (\tau )~\delta ( X^0(\tau ) - t)
~\delta ({{\bf X}} (\tau ) - {\bf r} )
\label{na2}
\end{equation}
is covariantly conserved, simply as a consequence of its definition (\ref{na2}), and of the equations (\ref{na1}) satisfied by $X^\mu, \xdot^\mu \propto P^\mu,$ and ${\cal Q}_a$.

A convenient choice of parameterization is $X^0(\tau ) = \tau$, so that $J^\mu_a \equiv (\rho_a, \vJ_a)$ is given by
\beqar
\rho_a (t, {\bf r} )&=& {\cal Q}_a (t) ~\delta ( {{\bf X}}(t) - {\bf r} ),\label{na3}\\
{{\bf J}}_a (t, {\bf r} )&=&  {\cal Q}_a (t) ~ {\dot {{\bf X}}}(t) ~\delta ({{\bf X}}(t)
- {\bf r} ).\label{na4}
\eeqar

To arrive at the corresponding fluid equations, we first consider an assembly of $N$ particles; {\it i.e.} we decorate all the co-moving variables with a particle-counting index $n= (1, ..., N): X^\mu_n, P^\mu_n, {\cal Q}_{an}$, and replace the single-particle formulas (\ref{na3}) and (\ref{na4}) with sums over n. Then we pass to a continuum/fluid description by letting the discrete counting variable $n$ become the continuous label $\bf x$, and we replace sums over $n$ by integrals over $\bf x$.
\beqar
\rho_a (t, {\bf r} ) &=& \int d^3x ~{\cal Q}_a (t, {\bf x} ) ~\delta ({{\bf X}}(t, {\bf x} ) - {\bf r}
),\label{na5}\\
{{\bf J}}_a(t,{\bf r} ) &=&\int d^3x ~ {\cal Q}_a (t,{\bf x} ) ~{\dot {{\bf X}}}(t, {\bf x} )
~\delta ({{\bf X}}(t, {\bf x} )- {\bf r} ) \label{na6}
\eeqar
The continuum version of (\ref{na1}) becomes (we suppress $m$ and $g$)
\begin{subequations}
\begin{eqnarray}
\xdddot_\mu (t, {\bf x}) &=& {\cal Q}_a (t, {\bf x}) F^{a}_{\mu \nu} \big(t, \vX (t, {\bf x})\big)  \xdot^\nu (t, {\bf x}),\label{na7a}\\
\dot{\cal Q}_a (t, {\bf x}) &=&- f_{abc} \bigg[A^b_o \big(t, \vX (t, {\bf x})\big) + \dot{\bf X} (t, {\bf x}) \cdot \vA^b \big(t, \vX (t, {\bf x})\big)\bigg] {\cal Q}_c (t, {\bf x}). \label{na7b}
\end{eqnarray}
\end{subequations}
As a consequence of these, $J^\mu_a$ of  (\ref{na5}), (\ref{na6}) is covariantly conserved.

Let us now observe that (\ref{na5}) and (\ref{na6}) describe an Eckart factorization for the non-Abelian case in the following manner.
In the course of the x-integration, the $\delta$-function evaluates $\bf x$ at function $\vec\chi \tbr$, such that
\begin{equation}
\vX \big(t, \vec\chi \tbr \big) = \bf r.
\label{na8}
\end{equation}
Also there is a Jacobian factor. Thus
\begin{equation}
\rho_a \tbr = Q_a \tbr \rho \tbr,
\label{na9}
\end{equation}
where
\begin{eqnarray}
Q_a (t, {\bf r} ) &\equiv& {\cal Q}_a (t, {\bf x} ) \bigg\vert_{{\bf x} = \chi \tbr ,}\label{na10}\\ [1ex]
\rho \tbr &\equiv& 1/\det \frac{\partial X^i (t, {\bf x})}{\partial x^j} \bigg\vert_{{\bf x} = \chi \tbr ,} \label{na11}\\
{{\bf J}}_a (t, {\bf r} ) &=& Q_a (t, {\bf r} ) ~{{\bf v}}(t, {\bf r}) ~ \rho (t, {\bf r} ),
\label{na12}\\
{\bf v} \tbr &\equiv& \dot{\bf X} (t, {\bf x}) \bigg\vert_{{\bf x} = \chi \tbr}.
\label{na13}
\end{eqnarray}
Thus we may write
\begin{eqnarray}
J^\mu_a (t, {\bf r} ) &=& Q_a (t, {\bf r} ) ~ j^\mu (t, {\bf r} ),
\label{na14}\\
j^\mu \tbr &=& \Big(\rho \tbr , {\bf v}\tbr  \rho\tbr \Big).
\label{na15}
\end{eqnarray}
Since $\rho \text{and}\ \bf j = \vv \rho$ are defined by integrals (\ref{na5}) and (\ref{na6}) without the charge ${\cal Q}_a  (t, {\bf x})$ factor, the Abelian current $j^\mu$ is ordinarily conserved.
\begin{equation}
\partial_\mu j^\mu = 0
\label{na16}
\end{equation}
Together with the covariant conservation of $J^\mu_a$, this implies that
\begin{equation}
j^\mu (D_\mu Q)^a =0,
\label{na17}
\end{equation}
which we call the fluid Wong equation. Note also that the Abelian current $j^\mu$ can be factorized {\it \'{a} la} Eckart as in (\ref{eqsix}), so that the complete non-Abelian current reads
\begin{equation}
J^\mu_a = Q_a nu^\mu, \qquad u^\mu u_\mu = 1.
\label{na18}
\end{equation}
[Unlike the Eckart decomposition for the Abelian current, (\ref{eqsix}), which entails no restriction, the non-Abelian decomposition (\ref{na18}) is specific to our assumption that a particle substratum underlies the fluid variables. An alternate picture is presented in Section 4.]

These results indicate that the color degree of freedom is carried by a color density $Q_a$, but all colored components move with the same velocity $\bf v$ or $u^\mu$ -- a single flow characterizes the fluid.

It remains to posit an  Euler force equation,  like (\ref{eqeight}), which captures the dynamics. This we do by specifying a Lagrangian action, which will also lead to (\ref{na16}) and (\ref{na17}). Moreover our Lagrangian will offer the possibility of generalizing the above single flow description to allow for different flows in  different components of the fluid \cite{bib4}, \cite{bib7}.

\section{Constructing the action}
The equations of motion for the non-Abelian fluid in the Euler formulation include
the kinematical equations:  continuity (\ref{eqeleven}) and Wong (\ref{na17}) that are
satisfied by the non-Abelian current, which is factorized as in (\ref{na18}). Still
needed is the non-Abelian Euler force equation, which extends (\ref{eqeight}),  or non-relativistically, (\ref{eqtwo}), thereby specifying the dynamics.
We present this by constructing an action whose vanishing variations reproduce the kinematical
equations and give a model for the dynamical equation. 

The algebra underlying the
non-Abelian theory is realized with anti-Hermitian generators $T^a$ satisfying
\beq
[T^a, ~T^b ] = f_{abc}~T^c,
\label{three1}
\eeq
and normalized by
\beq
{\rm tr }~ T^a T^b = - {1\over 2} \, \delta^{ab}.
\label{three2}
\eeq
In a canonical particle theory we expect that the algebra
(\ref{three1}) is reproduced by Poisson
brackets for corresponding symmetry generators.
In a field theory, we expect to find a copy of (\ref{three1}) at each point in space,
leading to the Poisson brackets
\beq
\{ \rho_a({\bf r}),~ \rho_b({\bf r}') \}
	= f_{abc}  ~\rho_c({\bf r} ) ~\delta({\bf r}-{\bf r}').
\label{three3}
\eeq
[A common time argument is suppressed.]
Upon quantization the brackets become commutators and 
acquire the factor $i /\hbar$.
(Schwinger terms do not spoil the quantum algebra
unless there are anomalies in the gauge symmetry.
Of course, we assume that the theory is anomaly-free.) \cite{bib9}

The action which leads to the bracketing rules (\ref{three3}) is the field theoretic version of the
Kirillov-Kostant form, which for a particle (not a field) reads \cite{bib10}
\beq
I_{K K} = 2 n \int dt~ tr T^3 g^{-1} {\dot g},
\label{three4}
\eeq
where $n$ is a normalization constant and we have taken the group $SU(2)$ as a concrete example: $g
\in SU(2), g= exp (T^a \varphi_a), T^a = \sigma^a/2i$ and
$\sigma^a$ are Pauli matrices. Also one uses (\ref{three4}) as part of an action for the Wong equation of motion for a particle.

More generally, consider a Lie group $G$ with $H$ denoting its Cartan subgroup (or
maximal torus).
The required generalized action is given by \cite{bib4}, \cite{bib7}
\beq
I_0 =  \sum^r_{s=1}~{\bar{n}_s} \int dt~ tr K_{(s)} g^{-1} {\dot g},
\label{three5}
\eeq
where $g\in G$ and $\bar{n}_s$ are the highest weights defining a unitary irreducible
representation of $G$, $K_{(s)}$ are the diagonal generators of the
Cartan subalgebra $H$ of
$G$. The summation in (\ref{three5}) extends up to the rank $r$ of the
algebra, but some of the $\bar{n}$'s could vanish.

Given this structure, we see that field theoretic generalization, which would give rise to (\ref{three3}), appears
as  $\int dt\, dr\, \sum ^r_{s=1} \bar{n}_{s} \, tr \, K_{(s)} \, g^{-1} \, \dot{g}$,
where now $\bar{n}_s$ and $g$ depend on $\bf r$, while the $K_{(s)}$
remain as constant elements of the Cartan subalgebra of the
group. Thus we take as the Lagrange density for our
non-Abelian fluid dynamics the formula \cite{bib4}, \cite{bib7}
\beq
{\cal L}= \sum_{s=1}^{r} j^\mu_{(s)}
2 ~\trrr  K_{(s)} g^{-1}D_\mu g
-f(n_{(1)},n_{(2)},\ldots, n_{(r)}) + {\cal L}_{gauge}.
\label{three6}
\eeq
Here $j^\mu_{(s)}$ are a set of Abelian currents; they may be
taken to be in the Eckart form
$j^\mu_{(s)} = n_{(s)} u^\mu_{(s)}$,
where $u^\mu$'s are four-velocity vectors, with
$	u^\mu_{(s)} u_{(s) \mu} = 1 $; ${\cal L}_{gauge}$ governs the gauge dynamics of the fields to which the fluid couples.
As far as the variational problem of this action is concerned, we regard
$n_{(s)}$ as given by $j^\mu_{(s)}$ via $n_{(s)} = \sqrt{ j^\mu_{(s)} j_{\mu (s)}}$.
The space and time components of the currents are given by
$j^\mu_{(s)}= (\rho_{(s)} , \rho_{(s)} {\bf v}_{(s)} )$, $\rho_{(s)} = n_{(s)}
u^0_{(s)}$.
In equation (\ref{three6})
$n_{(s)}$ are the invariant densities for the diagonal directions
of the Lie algebra.

Comparison with the usual form of the action for an Abelian fluid shows that what we
have obtained is the non-Abelian analogue of the irrotational part of the
flow. In the Abelian case, and without the gauge field coupling,
equation (\ref{three6}) entails a single contribution, $s=1$, and
$2\trrr (K_1 g^{-1} \del_\mu g) = - \del_\mu \theta$, with vanishing vorticity.
In the non-Abelian
case, the vorticity is nonvanishing.
One can nevertheless generalize (\ref{three6}) to
include the other components of the Clebsch parametrized vector (\ref{eqten}) which
couples to
$j^\mu_{(s)}$. This gives the Lagrange density \cite{bib4}, \cite{bib7}
\begin{equation}
\La = \sum\limits_{s=1}^{r} j^\mu_{(s)} \left\{
2 \trrr  K_{(s)}
g^{-1}D_\mu g +a_{\mu (s)}\right\}
    -f(n_{(1)},n_{(2)},\ldots, n_{(r)}) + \La_{gauge},
\label{three7}
\end{equation}
where $a_{\mu (s)}$ is given by
\beq
a_{\mu (s)} = \alpha_{(s)}\partial_\mu \beta_{(s)}.\label{three8}
\eeq
For the rank - one group $SU(2)$, with its single Cartan
element, the $s$-sum in (\ref{three6}) is exhausted by a single
element. We see that the $SU(2)$ fluid has one component.
For higher rank groups, the non-Abelian fluid can have up to
$r$ components, but fewer---indeed even just a single flow---
are possible when some of the densities
$n_{(s)}$ vanish. Mathematically, single-component fluids are the simplest, but
physically it is unclear what kinematical regimes of a quark-gluon plasma, for
example, would admit or even require such a reduction in flows.

The covariant derivative in (\ref{three6}), namely,
\beq
D_\mu g=\d_\mu g + A_\mu g, \label{three9}
\eeq
involves a dynamical non-Abelian gauge potential $A_\mu=A_\mu^a
T^a$ whose dynamics is provided by $\La_{gauge}$.
The first term in $\La$
contains the canonical 1-form for the fluid variables in our theory and determines their
symplectic structure
and canonical brackets. We have added the Hamiltonian
density part, the function $f (n_{(1)},n_{(2)},\ldots, n_{(r)})$,
which
describes the fluid dynamics. The theory is invariant under gauge transformations
with group element $U$,
\beq
\begin{array}{rcl}
g &\to& U^{-1} g \\
A_\mu &\to& U^{-1}\left(A_\mu + \d_\mu\right) U\;.
\end{array}
\label{three10}
\eeq
where $U\in G$. One can show that the canonical structure of our theory leads to the
charge-density algebra (\ref{three3}) \cite{bib4}, \cite{bib7}.

The current that couples to the potential $A_\mu$ in (\ref{three7}) reads (matrix rotation)
\beq
J^\mu=\sum\limits_{s=1}^{r} Q_{(s)} j^\mu_{(s)} \;,
\qquad \mbox{with}\qquad
Q_{(s)} = g K_{(s)} g^{-1}\;.
\label{three11}
\eeq
Invariance against arbitrary variation of $g$ ensures that (\ref{three11}) is
covariantly conserved as in (\ref{eqeleven}), $D_\mu J^\mu =0$, \cite{bib4}, \cite{bib7}
but we also need the conservation of
individual $j^\mu_{(s)}$ \cite{bib4}, \cite{bib7}.  This is achieved by considering special
variations of $g$ of the form
$\delta_{(s)} g = g K_{(s)} \lambda_{(s)}$.
These variations of $g$ lead to
\beq
\d_\mu j^\mu_{(s)} = \dot{\rho}_{(s)} + \nabla\cdot ({\bf v}_{(s)} \rho_{(s)}
)=0.
\label{three12}
\eeq
The fluid Wong equation which follows from the conservation of the
Abelian and non-Abelian currents now reads
\beq
\sum\limits_{s=1}^{r} j^\mu_{(s)}D_\mu Q_{(s)}=0 \;.
\label{three13}
\eeq

Varying the individual $j^\mu_{(s)}$ in (\ref{three7}) (recall that $n_{(s)} = \sqrt{j^\mu_{(s)} j_\mu (s)}$)
produces the non-Abelian  Bernoulli
equations
\beq
2 \trrr  Q_{(s)} (D_\mu g) g^{-1} = u_\mu\,
f^{(s)},
\label{three14}
\eeq
where
\beq
 f^{(s)} \equiv
	\frac{\d}{\d n_{(s)} } f(n_{(1)},n_{(2)},\ldots,n_{(r)}).
\label{three15}
\eeq
The curl of equation (\ref{three14}) can be cast in the
form
\beq
 \d^\mu \left(u^\nu_{(s)} f^{(s)}\right)
      - \d^\nu \left(u^\mu_{(s)} f^{(s)}\right)
=2 \trrr\bigg((D^\mu Q_{(s)}) (D^\nu g) g^{-1}
      +  Q_{(s)} F^{\mu\nu} \bigg) 
\;.
\label{three16}
\eeq
When contracted with $j^\mu_{(s)}=n_{(s)} u^\mu_{(s)}$, this leaves
\beq
n_{(s)}u^{\mu}_{(s)} 
	\d_\mu \left(u^\nu_{(s)} f^{(s)}\right)
      - n_{(s)} \d^\nu  f^{(s)}
=j_{\mu (s)} 2 \trrr \bigg( (D^\mu Q_{(s)}) (D^\nu g) g^{-1}
      + Q_{(s)} F^{\mu\nu} \bigg)
\;.
\label{three17}
\eeq
In the single channel case, the right side simplifies, since the sums in (\ref{three11}) and (\ref{three13}) contain only a single term. Suppressing the subscript (s), we are left from (\ref{three17}) with,
\beq
nu^{\mu}\,
	\d_\mu \left(u^\nu f^{\prime} (n)\right)
      - n \, \d^\nu  f^{\prime} (n)
=2 \trrr \, J_\mu\, F^{\mu\nu}
\;.
\label{three18}
\eeq
The left side is of the form of the usual Abelian Euler equation [compare (\ref{eqeight})]; the right side describes the non-Abelian Lorentz force acting on the charged fluid.

In the multichannel case, (\ref{three11}) and (\ref{three13}) are not available because no summation is performed on (\ref{three17}), and we must remain with that equation -- the non-Abelian Euler equation for multi-channel flow.

The matter part of the energy-momentum tensor is given as
\beq
\theta^{\mu\nu}=-g^{\mu\nu} \left(
	\sum\limits_{s=1}^{r} n_{(s)} f^{(s)} - f\right)
	+ \sum\limits_{s=1}^{r}
	u^\mu_{(s)} u^\nu_{(s)} \, n_{(s)} f^{(s)}\;.
\label{three19}
\eeq
Its divergence
\begin{subequations}\label{three20}
\beq
\d_\mu \theta^{\mu\nu} =  \sum\limits_{s=1}^{r}\left\{
	\d_\mu (n_{(s)}u^\mu_{(s)})
	\ u^\nu_{(s)} f^{(s)}
	+  n_{(s)}\left(u^\mu_{(s)} 
	\d_\mu\, (u^\nu_{(s)} f^{(s)})
	-  \d^\nu  f^{(s)}\right)
\right\}
\label{three20a}
\eeq
is evaluated with (\ref{three12}) and (\ref{three13}) as
\beq
\sum\limits_{s=1}^{r} j_{\mu (s)} 2 \trrr 
	\bigg( ( D^\mu Q_{(s)}) (D^\nu g) g^{-1}
	+  Q_{(s)} F^{\mu\nu}\bigg) \label{three20b}
.
\eeq
Since now we are summing over all channels, it follows from
(\ref{three11}) and (\ref{three17}) that,
\beq
\d_\mu \theta^{\mu\nu} =   2 \trrr J_\mu F^{\mu\nu} \, ,
\label{three20c}
\eeq
\end{subequations}
which is canceled by the divergence of the gauge-field
energy-momentum tensor.
\beq
\d_\mu \theta^{\mu\nu}_{gauge}= - 2 \trrr
	 J_\mu F^{\mu\nu} \label{three21}
\eeq
Thus we have conservation of the total energy-momentum tensor.

Some simplifications, which lead to a more transparent physical
picture, occur if
the dynamical potential separates.
\beqar
f(n_{(1)},\ldots,n_{(r)}) &=& \sum\limits_{s=1}^{r} f_{(s)}(n_{(s)})\nonumber\\
f^{(s)}&=& f'_{(s)}
\label{three22}
\eeqar
Then the left side of (\ref{three17}) refers only to variables
labeled $s$, while the right side may be rewritten with the help of
the non-Abelian Wong equation (\ref{three13}) to give
\beq
n_{(s)} u^{\mu}_{(s)}\, \d_\mu \left(
	u^\nu_{(s)} f'_{(s)}\right) - n_{(s)} \d^\nu  f'_{(s)}
=2 \trrr \bigg(J_{\mu} F^{\mu\nu} 
 - \sum\limits_{s'\neq s}^{r} j_{\mu (s')}
      \left( Q_{(s')} F^{\mu\nu} + (D^\mu Q_{(s')}) (D^\nu g)
g^{-1}\right) \bigg) \;.
\label{three23}
\eeq
Thus in the addition to the Lorentz force, there are forces
arising from the other channels $s'\neq s$.

Note that for separated dynamics (\ref{three22}), the energy-momentum
tensor (\ref{three19}) also separates,
\beq
\theta^{\mu\nu}=\sum\limits_{s=1}^{r} \theta^{\mu\nu}_{(s)}
=\sum\limits_{s=1}^{r}\left\{ -g^{\mu\nu}
	\left( n_{(s)} f'_{(s)} - f_{(s)}(n_{(s)}) \right)
	+   u^\mu_{(s)} u^\nu_{(s)} 
	n_{(s)} f'_{(s)}\right\}\;.
\label{three24}
\eeq
but the divergence of each individual $\theta^{\mu\nu}_{(s)}$ does not
vanish. This is just as expected since energy can now be
exchanged
between the different channels and with the gauge field;
this is also evident
from the equation of motion (\ref{three23}). It is clear that this
fluid moves with $r$ different velocities ${\bf v}_{(s)}$.

The single-channel Euler equation (\ref{three18}) is expressed in
terms of physically relevant quantities (currents, chromomagnetic
fields); the many-channel equation (\ref{three17}) involves,
additionally, the gauge group element $g$.  One may simplify
that equation by going to special gauge, for example $g=I$,
so that the right side of (\ref{three17}) reduces to
\beq\
j^{\mu}_{(s)} 2 \trrr \bigg( (D_\mu Q_{(s)}) (D_\nu g) g^{-1}
      +  Q_{(s)} F_{\mu\nu} \bigg)
= j^\mu_{(s)} 2 \trrr  K_{(s)}  (\d_\mu A_\nu-\d_\nu A_\mu),
\label{na44}
\eeq
while the Wong equation (\ref{three13}) becomes
\beq
\sum\limits_{s=1}^{r} j^\mu_{(s)} [A_\mu,K_{(s)}]=0.
\label{na45}
\eeq
It is interesting that in this gauge the nonlinear terms in
$F_{\mu\nu}$ disappear.

Observe that the inclusion of the $\alpha, ~\beta$-components of the \Cpr
current, or the use of the Lagrangian (\ref{three7}), does not change the form of the
equations of motion for fluids when expressed in terms of the velocities
and densities. The expressions for these quantities in terms of
the group parameters  and in terms of $\alpha$, $\beta$ will, of course, be altered.

We record the nonrelativistic limit of the single-channel Euler equation
(\ref{three18}). (For this calculation we restore the velocity of light c, which had been scaled to unity.)
For small velocities, we may write
\beqar
n &\approx& \rho -\frac{1}{2 c^2}\rho {\bf v}^2, \nonumber\\
u^\mu &\approx& (c , {\bf v}).\label{na28}
\eeqar
Further, we take $f(n)$ to be of the form
\beq
f=n c^2 +V(n)\, .
\label{na29}
\eeq
With these simplifications, we find that the
nonrelativistic limit for the spatial component of (\ref{three18})
gives the Euler equation with a non-Abelian Lorentz force,
\beq
\dot{\bf v} + {\bf v}\cdot\nabla {\bf v} = \vec{force}
	+ Q_a \vec{E}^a + \frac{{\bf v}}{c} \times Q_a \vec{B}^a\, ,
\label{na30}
\eeq
where $\vec{force}$ is the pressure force coming from the potential
$V$ (and is therefore Abelian in nature), while non-Abelian force
terms involve the chromoelectric and chromomagnetic fields.
\beq
E^i_a = c F_{0i}^a \;, \qquad
B^i_a = -\frac{1}{2} \epsilon^{ijk} F^a_{jk}\label{na31}
\eeq
It is seen that the non-Abelian fluid moves effectively in a
single direction specified by $\vec{j}=\rho\bf v$. Nevertheless,
it experiences a non-Abelian Lorentz force.

\section{Field-Based Fluid Mechanics}
The theory that we have developed  arose from an underlying particle picture of matter. Even before committing to definite dynamics (single-channel or multi-channel), very general considerations on the particle substratum led to the non-Abelian Eckart parameterization (\ref{na18}) and the fluid Wong equation (\ref{na17}). For contrast, I now present a derivation of  different non-Abelian fluid equations, which arise from a field (as opposed to particle) description of the fundamental substratum. This is the Madelung representation for the Schr\"{o}dinger theory: in the Abelian case the resulting fluid equations are of the usual Eulerian form. (This is in keeping with the fact that the Abelian Eckart form is identically true; it requires no committment to a specific physical picture.) In the non-Abelian case the equations differ from what has been previously derived within a particle description of the underlying matter.

A field-based realization of the Euler equations for an Abelian fluid is provided by
the Madelung ``hydrodynamical" rewriting of the Schr\"odinger equation \cite{bib11}. 
\beq
i\hbar \, \dot{\psi}
= -{\hbar^2 \over 2m} \vec \nabla^2 \psi
\label{four1}
\eeq
(We consider only the free equation.)
Upon presenting the wave function as
\beq
\psi (t, {\bf r} ) = \sqrt{\rho (t, {\bf r} )}~ e^{im \theta (t,{\bf r})/\hbar},
\label{four2}
\eeq
we find that the imaginary part of (\ref{four1}) results in the continuity equation
for
$j^\mu = (\rho , {\bf j})$, where the spatial current ${\bf j}$
is also the quantum current
$(\hbar /m)~{\rm Im} \psi^* \vec \nabla \psi$.
When ${\bf j}$ is written as $ {{\bf v}} \rho$, ${{\bf v}}$ is identified as
$\nabla \theta$; the velocity is irrotational.
The real part of (\ref{four1}) gives the Bernoulli equation, with a quantum force
derived from
\beq
V = {\hbar^2 \over 2m^2} (\vec \nabla \sqrt{\rho} )^2
= {\hbar^2 \over 8m^2} {(\vec \nabla \rho )^2 \over \rho}.
\label{four3}
\eeq
The Euler equation follows by taking the gradient of the Bernoulli equation.
In this way, we arrive again at the conventional irrotational fluid.

The story changes if we start from a non-Abelian Schr\"odinger equation.
Again we consider the free case with nonrelativistic kinematics.
Thus the equation involves a multi-component wave function
$\Psi$, with
\beq
i\hbar \dot{\Psi}
= -{\hbar^2 \over 2m} \vec \nabla^2 \Psi.
\label{four4}
\eeq
The color degrees of freedom lead to the conserved non-Abelian current.
\beqar
J^\mu_a &=& (\rho_a , {{\bf J}}_a ) \nonumber\\
\rho_a = i \Psi^\dagger T^a \Psi,  && ~~~~
{{\bf J}}_a = {\hbar \over m} {\rm Re} \Psi^\dagger T^a \vec \nabla \Psi
\label{four5}
\eeqar
The singlet current $j^\mu = (\rho , {\bf j})$ is also conserved.
For definiteness and simplicity, we shall henceforth
assume that the group is $SU(2)$ and that the representation is the
fundamental one: $T^a={\sigma^a}/{2i}$, $\{T^a,T^b\}=-\delta_{ab}/2$.
We shall also set the mass $m$ and Planck's constant $\hbar$ to unity. The
non-Abelian analogue of the Madelung decomposition
(\ref{four2}) is \cite{bib4}, \cite{bib7}, \cite{bib12}
\beq\label{four6}
\Psi=\sqrt{\rho} ~g u\, ,
\eeq
where $\rho$ is the scalar $\Psi^\dagger\Psi$, $g$ is a
group element, and $u$ is a constant vector that points in
a fixed direction [{\it e.g.}, for $SU(2)$ the two-component
spinor $u$ could be taken as $u_1 =1,~u_2 =0$, then $i
u^\dagger T^a u= \delta^a_{3}/2$].  The singlet density is $\rho$,
while the singlet current ${\bf j} = I m \Psi^\ast \vec \nabla \Psi$ is
\beq\label{four7}
{\bf j}= {\bf v}\rho\;, \qquad {\bf v}\equiv
	-i u^\dagger  \:g^{-1} \vec\nabla g\:u
\;.
\eeq
With the decomposition (\ref{four6}), the color density (\ref{four7})
becomes
\beq\label{four8}
\rho_a=Q_a \rho\;,\qquad Q_a= i u^\dagger\: g^{-1}T^a g \: u = i R_{ab}~
u^\dagger T^b u =\ R_{ab}~ t^b/2,
\eeq
where $R_{ab}$ is in the adjoint representation of the group and
the unit vector $t^a$ is defined as $t^a/2= i u^\dagger T^a u$. On
the other hand, the color current reads
\beq\label{four9}
{\bf J}_a =\frac{1}{2} \rho~ R_{ab} ~u^\dagger \:(T^b
	\: g^{-1}\vec \nabla g + g^{-1}\vec \nabla g \: T^b) \: u \;,
\eeq
which with the introduction of
\beqar
g^{-1}\nabla g &\equiv& -2{\bf v}^a T^a\, ,\label{four10}\\
{\bf v}&=&{\bf v}^a t^a\, ,\label{four11}
\eeqar
may be presented as
\beq
{\bf J_a}=\frac{\rho}{2} R_{ab} ~{\bf v}^b\;.
\label{four12}
\eeq
Unlike the Abelian model, the vorticity is nonvanishing.
\beq
\vec \nabla\times{\bf v}^a = \epsilon^{abc} {\bf v}^b\times {\bf v}^c
\label{four13}
\eeq

A difference between the Madelung approach and the previous
particle based one is that the color current is not proportional
to the singlet current, as in the non-Abelian Eckart parameterization. Equation (\ref{four12}) may be decomposed as
\beq
{\bf J}_a =  Q_{a}\rho {\bf v}
	+\frac{\rho}{2} R_{ab} {\bf v}^b_\perp
\label{four14}
\eeq
where the ``orthogonal'' velocity ${\bf v}_\perp^a$ is defined as
\beq
{\bf v}^a_\perp=\left(\delta^{ab}-t^a t^b\right) {\bf v}^b\;.
\label{four15}
\eeq
Equation (\ref{four14}) shows that color current possesses
components that are orthogonal to the singlet current.

The decomposition of the SU(2) Schr\"odinger equation with the
parametrization (\ref{four6}) results in two equations: one regains
the conservation of the Abelian current and the
other is the ``Bernoulli'' equation.
\beq
\left(  g^{-1} \dot{g}\right)^a = \left[ {\bf v}^b\cdot {\bf v}^b
	- \frac{\vec \nabla^2 \sqrt{\rho}}{\sqrt{\rho}}\right] t^a
	+ \frac{1}{\rho} \vec \nabla\cdot \left(\rho \epsilon_{abc}
	{\bf v}^b t^{c}\right)
\label{four16}
\eeq
It is further verified that the covariant conservation
of the color current is a consequence of the Abelian continuity equation
 and of
(\ref{four16}). However, there is no Wong equation because the
color current is not proportional to the conserved singlet
current. Finally, using the identity, which follows from
the definition (\ref{four10}),
\beq
\dot{\bf v}^a = -\frac{1}{2}\vec \nabla \left( g^{-1} \dot{g}
\right)^a + \epsilon_{abc} {\bf v}^b \left(g^{-1} \dot{g}
\right)^c\, , \label{four17}
\eeq
one can deduce an Euler equation for $\dot{\bf v}^a$ from
(\ref{four16}). No equation emerges that is analogous to the non-Abelian fluid equation (\ref{na17}).

It is clear that within
the present approach the fluid color flows in every direction in
group space, but the mass density is carried by the unique
velocity ${\bf v}$.  This is in contrast to our previous approach
where all motion is in a single direction
or at most in the directions of the Cartan elements of the Lie
algebra.
The difference between the two approaches is best seen from a
comparison of Lagrangians. For the color Schr\"odinger theory in
the Madelung representation
\beqar
\La_{Schrodinger}&=&\frac{i}{2}\left(\Psi^\dagger\dot{\Psi}
	- \dot{\Psi}^\dagger \:\Psi \right)
	-\frac{1}{2} \vec \nabla\Psi^\dagger\cdot \vec \nabla \Psi \label{four18}
\\
&=& i\rho \:u^\dagger \:g^{-1}\dot{g}\: u
	-\frac{1}{2}\rho {\bf v}^a\cdot {\bf v}^a
	-\frac{(\vec \nabla\rho)^2}{8 \rho}.\label{four19}
\eeqar
With $u\otimes u^\dagger \equiv I/2 - 2 i K$, the free part of
the above reads
\beq
\La^0_{Schrodinger} = \rho \: 2\trrr K g^{-1}\dot{g}
-\frac{1}{2} \rho {\bf v}^a\cdot {\bf v}^a\, .
\label{four20}
\eeq
On the other hand, the free part of the Lagrange density
(\ref{three6})
in the nonrelativistic limit, with $f(n)$ given by $nc^2 = \sqrt{\rho^2 (c^2 - {\bf v}^2)}$ (we restore the velocity of light c), reads
\beqar
\La^0&=&\rho \:2\trrr\bigg( K g^{-1}\dot{g} 
	+ K {\bf v} \cdot g^{-1}\vec \nabla g\bigg)
	- \sqrt{\rho^2 (c^2-{\bf v}^2)}\, ,
\nonumber \\
&\approx& \rho \:2\trrr\, \bigg( K g^{-1}\dot{g} 
	+ K {\bf v} \cdot g^{-1}\vec \nabla g \bigg)
	- \rho c^2+\frac{1}{2}\rho {\bf v}^2\, ,
\label{four21}\\
&=& \rho \:2\trrr\,  K g^{-1}\dot{g} 
	-\frac{1}{2}\rho {\bf v}^2 - \rho c^2\, , \nonumber
\eeqar
where we have used ${\bf v}=-2 \trrr\, K g^{-1} \nabla g$,
which follows upon the variation of ${\bf v}$, in the next-to-last
equality above. Thus the canonical $1$-form is the same for both
models while the difference resides in the velocity dependence
of their respective Hamiltonians.  Only the singlet ${\bf v}$
enters (\ref{four21}) while the Madelung construction uses the
group vector ${\bf v}^a$.
\par
Finally, note that while the Euler equation, which emerges when
(\ref{four16}) and (\ref{four17}) are combined, intricately couples
all directions of the fluid velocity ${\bf v}^a$, it does admit
the simple solution ${\bf v}^a = {\bf v}t^a$, with ${\bf v}$ obeying
the Abelian equations that arise from (\ref{four1})-(\ref{four2}).

\end{document}